\documentclass[preprint,showpacs,preprintnumbers,amsmath,amssymb]{revtex4}

\usepackage{graphicx}
\usepackage{dcolumn}
\usepackage{bm}

\newcommand{\nab}{{\bf{\nabla}}}

\newcommand{\lam}{\Lambda(t)}

\newcommand{\pa}{\partial}

\newcommand{\largep}{{\bf p}}

\newcommand{\ep}{\varepsilon}

\newcommand{\larger}{{\bf r}}

\newcommand{\setr}{\{\larger\}}
\newcommand{\hanaV}{\mathcal{V}_{FF}}

\newcommand{\sumj}{\sum_{j=1}^N}
\newcommand{\re}{\mbox{Re}}
\newcommand{\im}{\mbox{Im}}
\newcommand{\vap}{\varphi}

\begin{document}
\title{Acceleration of adiabatic transport of interacting particles
and rapid manipulations of dilute Bose gas in ground state}
\author{Shumpei Masuda}
\email[]{syunpei710@cmpt.phys.tohoku.ac.jp}
\affiliation{
Department of Physics,
Tohoku University,
Sendai 980, Japan}
\date{\today}
\begin{abstract}
We show a method to accelerate quantum adiabatic transport of 
identical spinless particles interacting with each other 
by developing the preceding fast-forward scaling theory formed for 
one-particle systems
[Masuda and Nakamura, {\it Proc. R. Soc.} A {\bf 466\/}, 1135 (2010)].
We derive a driving potential which accelerates adiabatic dynamics of 
quantum systems composed of identical particles
in order to obtain the final adiabatic states in any desired short time. 
We also exhibit an ideal rapid manipulation of dilute 
Bose gas in the ground state without energy excitation 
by using the fast-forward scaling theory.
\end{abstract}

\maketitle

\section{Introduction}
Technology to manipulate quantum states is rapidly evolving, and 
various methods to control quantum states have been reported in
Bose Einstein condensates (BEC)\cite{leg,gus,ket,lea},
in quantum computing \cite{nie} and in many
other fields of applied physics.  
It would be very important to consider the acceleration of such
manipulations of quantum states for manufacturing purposes and 
for innovation of technologies.
Earlier we proposed the acceleration of quantum
dynamics  \cite{mas1} and quantum adiabatic dynamics \cite{mas2,mas3}. 
The theory is called “fast-forward
scaling theory" or "fast-forward theory". 
We derived a driving potential which accelerates given
quantum dynamics and generates exactly a target states 
in any desired short time,
where the target state is defined as the final state in the given original 
dynamics. 

The acceleration of quantum adiabatic dynamics is very 
important for many current and future technologies. 
Adiabatic manipulations seem to be an ideal method for the control 
of quantum systems because of the adiabatic theorem \cite{kat}.
However adiabatic dynamics can take too long time compared with 
the lifetime or coherent time of the system \cite{Zur}.
Acceleration of adiabatic dynamics overcomes the difficulty.  
Various methods of acceleration of adiabatic dynamics 
or shortcut to adiabaticity have been proposed: 
fast-forward scaling theory \cite{mas2,mas3},
frictionless quantum driving \cite{ber3} and
invariant-based inverse engineering\cite{mug1}. 
Recently applications of these methods  
to the control of BEC have been proposed 
theoretically \cite{mas2,che,mug2,torr,Cam}, and been
demonstrated experimentally \cite{Scha,Scha2,Bas}.
However, the range of
applications of the methods is still limited in simple cases.
Construction of the theory for many-body systems and for more 
general controls is important and useful for various kinds of 
manipulations of quantum systems. 


In this paper we extend the previous scheme of the acceleration of adiabatic 
dynamics to many-body systems.
Our theory combines opposite ideas: the infinitely fast acceleration and 
ultimately slow adiabatic dynamics, and
uses a space-dependent additional phase to give the driving potential. 
We exhibit acceleration of adiabatic transport of identical spinless 
particles interacting with each other.
We show a driving potential which conveys the interacting 
particles without energy excitation.
We also propose an ideal rapid 
manipulation of dilute Bose gas in the ground state.
By using the method, the final state of the Bose gas in the original adiabatic
dynamics is generated in any short time without energy excitation.
In Section \ref{Extension of the formalizm of fast-forward to many-body 
systems} we derive a formula of the driving potential
for many-body system composed of identical spinless particles.
In Section \ref{Examples} we exhibit
acceleration of adiabatic transport of interacting particles
and rapid manipulations of dilute Bose gas by using fast-forward scaling theory.
Section \ref{Conclusion}. is devoted to conclusion.

\section{Fast-forward theory in many-body systems}
\label{Extension of the formalizm of fast-forward to many-body 
systems}
We extend the framework of the fast-forward theory 
formed for one-particle systems to many-body systems 
composed of identical spinless particles.
We derive a driving potential which realizes the final state of a 
given original dynamics from its initial state.
First we shall derive a formula of the driving potential 
for (non-adiabatic) standard 
dynamics as a preparation for the acceleration of adiabatic dynamics.
The formula is used in the derivation of a 
driving potential for adiabatic dynamics.
\subsection{Standard fast-forward}
We consider a system composed of $N$ identical spin-less particles interacting 
with each other.
Hamiltonian is given by
\begin{eqnarray}
H_0 = \sum_{j=1}^N\frac{\largep_j^2}{2m} + V_{e}(\{\larger\},t)+V_I(\{\larger\}),
\label{h01}
\end{eqnarray} 
where $j$ denotes the particles,
and $m$ is mass.
$\{\larger\}=(\larger_1, \larger_2, \cdots , \larger_N)$ 
denotes a set of coordinates of all the particles.
$V_e$ is an external potential expressed by one-particle operators
$v_e(\larger,t)$ as
\begin{eqnarray}
V_{e}(\{\larger\},t)=\sum_{j=1}^N v_e(\larger_j,t).
\end{eqnarray} 
$V_I$ is a time-independent interaction potential which is a
function of the relative positions of the particles.
$\Psi_0=\Psi_0(\{\larger\},t)$ is a many-body wave function 
ruled by $H_0$.
We call $\Psi_0$ standard state.
Instead of a simply accelerated state of $\Psi_0$
we consider the fast-forwarded state $\Psi_{FF}$ defined by
\begin{eqnarray}
\Psi_{FF}(\{\larger\},t)= \Psi_0(\{\larger\},\lam)e^{if(\{\larger\},t)},
\label{pf1}
\end{eqnarray}
with the additional phase $f(\{\larger\},t) \in R$,
because it is not possible to realize the simply accelerated state
$\Psi_0(\{\larger\},\lam)$ \cite{mas1}.
$\lam$ is defined as
\begin{eqnarray}
\lam = \int_0^t\alpha(t')dt',
\label{lam_8}
\end{eqnarray}
where
$\alpha(t)\in R$ is the magnification factor of the fast-forward
which characterizes the intensity of the acceleration.
Time-dependence of $\alpha$ is tuned so that the additional phase $f$
disappears at the initial and final time of the 
acceleration (The detail is shown
later).
We define $\Psi_0$ at time $T$ as the target state.
And arbitrary time $T_F>0$ is the final time of the acceleration.
$\alpha$ relates $T_F$ and $T$ through
\begin{eqnarray}
T = \int_0^{T_F}\alpha(t)dt.
\label{TandTF}
\end{eqnarray}
Driving Hamiltonian for $\Psi_{FF}$ is assumed as
\begin{eqnarray}
H_{FF}(\setr,t)=\sumj\frac{\largep_j^2}{2m} + V_e(\{\larger\},\lam) + 
V_I(\{\larger\}) + \mathcal{V}_{FF}(\{\larger\},t),
\label{hf1}
\end{eqnarray}
where $\hanaV$ is called driving potential.
Schr$\ddot{\mbox{o}}$dinger equation is represented as
\begin{eqnarray}
i\hbar\frac{\pa \Psi_{FF}}{\pa t}=H_{FF}\Psi_{FF}.
\label{sc1}
\end{eqnarray}

By using Eq.(\ref{pf1}) and 
Schr$\ddot{\mbox{o}}$dinger equations for $\Psi_0$ and $\Psi_{FF}$ we
can derive the equation
\begin{eqnarray}
-\hbar\frac{\pa f}{\pa t}|\Psi_0(\lam)|^2
&+&\Big{(}\alpha(t)-1\Big{)}\Big{[}
\sumj -\frac{\hbar^2}{2m}\Psi_0^\ast(\lam)\nab_j^2\Psi_0(\lam)
+\Big{(}V_e(\lam)+V_I\Big{)}|\Psi_0(\lam)|^2\Big{]}\nonumber\\
&+&\frac{\hbar^2}{2m}\sumj\Big\{ 2i\nab_jf\cdot\Psi_0^\ast(\lam)\nab_j
\Psi_0(\lam) + i(\nab_j^2 f)|\Psi_0|^2 -(\nab_j f)^2|\Psi_0(\lam)|^2
\Big\}\nonumber\\
&&=\hanaV(t) |\Psi_0(\lam)|^2
,\label{eq1_7_31}
\end{eqnarray}
where $f(\setr,t)$ is abbreviated by $f$.
By decomposing Eq.(\ref{eq1_7_31}) into real and imaginary parts, 
we can obtain the driving 
potential and the additional phase as
\begin{eqnarray}
\frac{\hanaV}{\hbar}&=&
-\frac{\pa f}{\pa t}
-\sumj\Big\{(\alpha-1)\frac{\hbar}{2m}\mbox{Re}\big{[}\nab_j^2\Psi_0/\Psi_0
\big{]}
\nonumber\\
&&+\frac{\hbar}{m}\nab_jf\cdot\mbox{Im}\big{[}\nab_j\Psi_0/\Psi_0\big{]}
+\frac{\hbar}{2m}(\nab_jf)^2\Big\}
+(\alpha-1)\frac{V_e+V_I}{\hbar}
\label{hanaV}
\end{eqnarray}
and
\begin{eqnarray}
f(\setr,t) = \big{(}\alpha(t)-1\big{)}\eta(\setr,\lam),
\label{f1}
\end{eqnarray}
respectively.
In Eq.(\ref{hanaV}) $f$, $\alpha$, $\Psi_0$, $V_e$ abbreviate $f(\setr,t)$,
$\alpha(t)$, $\Psi_0(\setr,\lam)$ and $V_e(\setr,\lam)$, respectively.
We suppose $\eta(\setr,t) \in R$ is the phase of $\Psi_0(\setr,t)$, $i.e.$,
$\Psi_0(\setr,t) = \tilde{\Psi}_0(\setr,t)\exp[i\eta(\setr,t)]$ where
$\tilde{\Psi}_0$ is the real amplitude of $\Psi_0$.
Because $f$ in Eq.(\ref{f1}) includes the factor $\alpha(t)-1$, 
the additional phase disappears everywhere when $\alpha = 1$.
Therefore we tune $\alpha$ so that $\alpha$ becomes unity 
at the initial and the final 
time of the acceleration and Eq.(\ref{TandTF}) is satisfied.
In Eq.(\ref{hanaV}), a space-independent term was neglected,
because it is concerned only with the space independent phase on $\Psi_{FF}$
and we are not concerned about it.

We should note that, although $\hanaV$ accelerates the dynamics, 
we can not generate $\hanaV$ in general
because we can not control general many-body potentials.
This scheme is basically applicable 
when $\hanaV$ in Eq.(\ref{hanaV}) is expressed in terms of one-body
operators $v_{FF}$ as 
\begin{eqnarray}
\mathcal{V}_{FF}= \sumj v_{FF}(\larger_j,t).
\label{hana_8_13_1}
\end{eqnarray}
In the next section such accelerations are shown after extending the
above formula for
acceleration of adiabatic dynamics in the following subsection.

\subsection{Acceleration of adiabatic dynamics}
We derive a driving potential which accelerates adiabatic dynamics
by using the formula in the previous subsection.
We consider adiabatic dynamics of a system composed of $N$ spin-less particles.
Hamiltonian is given by
\begin{eqnarray}
H_0 = \sumj\frac{\largep_j^2}{2m} + V_e(\{\larger\},R(t)) 
+ V_I(\{\larger\}).
\end{eqnarray} 
External field $V_e$ is a function of an adiabatic parameter $R$ defined by
\begin{eqnarray}
R(t) = R_0 + \ep t.
\label{R}
\end{eqnarray}
$R_0$ is the initial value of $R$.
The constant value $\ep$ is the rate of adiabatic change in $R(t)$. 
$\ep$ is infinitesimally small, $\ep \ll 1$.
$V_I$ is a time-independent interaction potential.
We suppose that the system is in $n$-th energy eigenstate of an instantaneous
Hamiltonian $H(R)$ in the adiabatic dynamics.
Wave function is represented as
\begin{eqnarray}
\Psi_0=\Psi_0(\{\larger\},t) = 
\phi_n(\{\larger\},R(t))e^{-\frac{i}{\hbar}\int_0^tE_n(R(t'))dt'}e^{i\gamma(t)},
\label{phin1}
\end{eqnarray} 
where $\phi_n(\{\larger\},R)$ is the $n$-th energy eigenstate
of the instantaneous Hamiltonian.
$E_n$ is the eigenenergy.
$\gamma(t)$ is an adiabatic phase.
$\phi_n$ satisfies 
\begin{eqnarray}
H_0(R)\phi_n(R) = E_n(R)\phi_n(R).
\label{sch1}
\end{eqnarray} 

Now we consider the acceleration of the adiabatic dynamics.
The wave function and Hamiltonian should be regularized \cite{mas2},
because we need a standard state which
satisfies Schr$\ddot{\mbox{o}}$dinger equation up to $O(\ep)$
to apply the theory developed in the previous subsection.
Standard wave function is modified with a phase $\ep\theta(\setr,t)$ as
\begin{eqnarray}
\Psi_0^{(reg)}(\{\larger\},R,t) = 
\phi_n(\{\larger\},R)e^{-\frac{i}{\hbar}\int_0^tE_n(R(t'))dt'}e^{i\ep\theta}.
\label{phi_reg}
\end{eqnarray} 
The regularized standard Hamiltonian is defined as
\begin{eqnarray}
H_0^{(reg)} = \sumj\frac{\largep_j^2}{2m} + V_e(\{\larger\},R(t)) 
+ V_I(\{\larger\}) + \ep \tilde{V}(\{\larger\},t).
\label{hreg1}
\end{eqnarray} 
$\theta=\theta(\setr,t)$ and $\tilde{V}=(\setr,t)$ are introduced so that 
the Schr$\ddot{\mbox{o}}$dinger equation:
\begin{eqnarray}
i\hbar\frac{\pa\Psi_0^{(reg)}}{\pa t}=H_0^{(reg)}\Psi_0^{(reg)}
\end{eqnarray} 
is satisfied up to $O(\ep)$.
Then $\theta$ should satisfy 
\begin{eqnarray}
\sumj\frac{\hbar}{2m}\Big{[}
\nab_j^2\theta + 2\mbox{Re}[\nab_j\phi_n/\phi_n]
\cdot\nab_j\theta\Big{]}
+\mbox{Re}\Big{[}\frac{\pa\phi_n}{\pa R}\Big{/}\phi_n\Big{]}
=0.\label{theta2}
\end{eqnarray}
And $\tilde{V}$ is given by
\begin{eqnarray}
\frac{\tilde{V}}{\hbar}=-\im\Big{[}\frac{\pa\phi_n}{\pa R}\Big{/}\phi_n\Big{]}
-\sumj\frac{\hbar}{m}\Big{[}\nab_j\theta\cdot\im[\nab_j\phi_n/\phi_n]\Big{]}.
\label{tV}
\end{eqnarray}
We define a fast-forwarded state and driving Hamiltonian 
with $\Psi_0^{(reg)}$ as
\begin{eqnarray}
\Psi_{FF}(\setr,t)&=&\Psi_0^{(reg)}(\setr,\lam)e^{if(\setr,t)}, \label{pff1} \\
H_{FF}(t)&=&\sumj\frac{\largep_j^2}{2m} + V_e(\{\larger\},\lam) + 
V_I(\{\larger\}) + \mathcal{V}_{FF}(\{\larger\},t),
\end{eqnarray}
where $f$ is the additional phase.
$\lam$ is defined by Eq.(\ref{lam_8}).
We take the limit: $\ep\rightarrow 0$, $\alpha = O(1/\ep) 
\rightarrow \infty$ and $\ep\alpha = O(1)$, that is,
we consider the infinitely fast acceleration of the
ultimately slow dynamics. 
$\alpha$ relates initial $(R_0)$ and final $(R_1)$ values of 
the adiabatic parameter
as
\begin{eqnarray}
R_1 - R_0 = \ep\int_0^{T_F}\alpha(t)dt,
\end{eqnarray}
where $T_F$ is the arbitrary final time of the acceleration.
By using Eq.(\ref{sch1}) in the Schr$\ddot{\mbox{o}}$dinger equation
of $\Psi_{FF}$ we can obtain the driving potential and the additional phase 
in an analogous manner to \cite{mas2} as
\begin{eqnarray}
\frac{\mathcal{V}_{FF}}{\hbar}
&=& 
-\frac{d\alpha}{dt}\ep\theta-\alpha^2\ep^2
\frac{\pa\theta}{\pa R}
-\sumj\frac{\hbar}{2m}\alpha^2\ep^2(\nab_j\theta)^2
\nonumber\\
&&-\alpha\ep\mbox{Im}\Big{[}\frac{\pa\phi_n}{\pa R}\Big{/}\phi_n\Big{]}
-\sumj\alpha\ep\frac{\hbar}{m}
\mbox{Im}\Big{[}\frac{\nab_j\phi_n}{\phi_n}\Big{]}\cdot\nab_j\theta,
\label{V1}
\end{eqnarray}
and
\begin{eqnarray}
f(\setr,t) &=& (\alpha-1)\ep\theta,
\label{f_8_16}
\end{eqnarray}
respectively.
In Eqs.(\ref{V1}) and (\ref{f_8_16}) 
$\alpha(t)$, $\theta\big{(}\setr,R(\lam)\big{)}$ and 
$\phi_n\big{(}\setr,R(\lam)\big{)}$ are abbreviated by 
$\alpha$, $\theta$ and $\phi_n$, respectively.
In Eq.(\ref{V1}) we omit the terms which are space-independent because they
concern only with space-independent phase.
$\ep\alpha$ is tuned to be zero at the initial and final time of the 
acceleration so that the additional phase in Eq.(\ref{f_8_16}) vanishes. 
It should be emphasized that the driving potential should be
one-particle potential, because we can not control
general many body potentials. 
Thus the driving potential should be written as
Eq.(\ref{hana_8_13_1}).
Moreover the additional phase must not change the statistics of the system.
In general, the driving potential in Eq.(\ref{V1}) 
can not be represented as Eq.(\ref{hana_8_13_1}) because it can contain
unrealizable many-body potential.
However in some specific but important manipulations, the driving potential
becomes a one-particle potential.
Such manipulations are exhibited in the following section.

\section{Applications}
\label{Examples}
\subsection{Acceleration of adiabatic transport of identical particles}
\label{Fast-forward transport of many body system}
We consider adiabatic
transport of identical particles trapped in an external potential.
The particles are conveyed into $x-$direction by a driving potential.
We assume that the system is in the $n$-th energy eigenstate of the
instantaneous Hamiltonian.
$\phi_n$ in Eq.(\ref{phi_reg}) is represented with $\phi'_n$ as
\begin{eqnarray}
\phi_n(\setr,R(t)) = \phi'_n(\{x-R(t),y,z\}),
\label{phin2}
\end{eqnarray}
where $\phi'_n(\setr)$ is a stationary wave function of 
$n$-th energy eigenstate trapped by a stationary potential $V'_e(\setr)$.
External potential $V_e$ in Eq.(\ref{hreg1}) is written
by $V'_e$ as
\begin{eqnarray}
V_e(\setr,R(t)) = V'_e(\{x-R(t),y,z\}).
\end{eqnarray}
The fast-forwarded state is defined by Eq.(\ref{pff1}) with $\phi_n$
in Eq.(\ref{phin2}).
The phase $\theta(\setr,t)$ of wave function in Eq.(\ref{phi_reg}) 
is a solution of Eq.(\ref{theta2}).
Noting that 
\begin{eqnarray}
\frac{\pa \phi_n}{\pa R}=-\sumj\frac{\pa\phi_n}{\pa x_j},
\end{eqnarray}
we obtain the solution as
\begin{eqnarray}
\theta = \sumj\frac{m}{\hbar}x_j.
\label{theta1}
\end{eqnarray}
On the other hand, it turns out that $\tilde{V}$ in Eq.(\ref{tV})
vanishes everywhere.
Substituting Eq.(\ref{theta1}) into Eq.(\ref{V1})
we obtain the driving potential as
\begin{eqnarray}
\mathcal{V}_{FF}=-\frac{d\alpha}{dt}\ep m\sumj x_j.
\end{eqnarray}
Therefore by applying the driving potential
\begin{eqnarray}
v_{FF}(\larger) = -\frac{d\alpha}{dt}\ep m x,
\label{vff3}
\end{eqnarray}
we can accelerate trapped particles without energy excitation at the final
time of the transport. 
The wave function 
of fast-forwarded state has the additional phase in Eq.(\ref{f_8_16})
during the transport.
The additional phase does not change the statistics of the system because 
$\theta$ is given by Eq.(\ref{theta1}).
$\alpha\ep$ is chosen to be zero at the initial and final time of the 
transport so that the additional phase vanishes and the fast-forwarded state
coincides with the target state.
Distance of the transport is given by $\ep\int_0^{T_F}\alpha(t)dt$.
Moreover, although we have considered  
energy eigenstates so far, it turns out from Eq.(\ref{vff3}) that
the driving potential can transport not only energy eigenstates but also
general states which is a superposition of energy eigenstates
because the driving potential does not depend on energy levels.

\subsection{Rapid manipulation of dilute Bose gas in ground state}
We propose an ideal rapid 
manipulation of dilute Bose gas in the ground state
by using fast-forward scaling theory.
Let us suppose that a system is composed of $N$ identical 
Bose particles with mass $m$.
We assume that the wave function of the ground state
is represented as
\begin{eqnarray}
\phi_n(\setr,R)=\vap(\larger_1,R)\vap(\larger_2,R)\cdots\vap(\larger_N,R)
\label{phi1}
\end{eqnarray}
for an instantaneous Hamiltonian $H(R)$, where
$\vap$ is a one-particle wave function parameterized by $R$
(mean field approximation \cite{Ued}).
In general $\vap$ is not the wave function of the ground state 
of an non-interacting boson.
We consider acceleration of the adiabatic dynamics of the Bose gas. 
Naive controls of external field without fast-forward theory would make 
the state much more complex than that in Eq.(\ref{phi1})
with energy excitation.
By using fast-forward theory we can 
transform the state from $\phi_n(\setr,R_{0})$
to $\phi_n(\setr,R_{1})$ in any short time, 
where $R_{0}$ and $R_{1}$ are the initial and 
final values of the adiabatic parameter.

Equation (\ref{theta2}) is rewritten 
by using Eq.(\ref{phi1}) as
\begin{eqnarray}
\sumj\Big\{\frac{\hbar}{2m}\Big{(}2\nab_j\theta\cdot\re [\nab_j\vap_j/\vap_j]
+\nab_j^2\theta\Big{)}+\re\Big{[}\frac{\pa \vap_j}{\pa R}\Big{/}
\vap_j\Big{]}\Big\}
=0,
\label{theta3}
\end{eqnarray}
where $\vap_j$ denotes $\vap(\larger_j,R)$.
In the derivation of Eq.(\ref{theta3}) we used 
\begin{eqnarray}
\frac{\pa\phi_n}{\pa R}\Big{/}\phi_n &=&
\sumj\frac{\pa\vap_j}{\pa R}\Big{/}\vap_j,
\label{eq1}\\
\nab_j\phi_n/\phi_n&=&\nab_j\vap_j/\vap_j.
\label{eq2}
\end{eqnarray}
We assume the form of $\theta$ as 
\begin{eqnarray}
\theta = \sumj a(\larger_j,R),
\label{a2}
\end{eqnarray}
where $a(\larger_j,R)$ is real.
Substituting Eq.(\ref{a2}) in Eq.(\ref{theta3}) it turns out that
$a_j$ is obtained by solving  
\begin{eqnarray}
\frac{\hbar}{2m}\Big{(}2\nab_ja_j\cdot\re [\nab_i\vap_j/\vap_j]
+\nab_j^2a_j\Big{)}+\re\Big{[}\frac{\pa \vap_j}{\pa R}\Big{/}\vap_j\Big{]}=0.
\label{a1}
\end{eqnarray}
Let us suppose that $\xi(\larger,R)$ is the phase of $\vap(\larger,R)$.
Then the phase $\eta(\{\larger\},R)$ of $\phi_n(\{\larger\},R)$ is written as 
\begin{eqnarray}
\eta(\{\larger\},R)=\sumj\xi(\larger_j,R).
\label{eq_8_21_1}
\end{eqnarray}
By using Eqs.(\ref{eq1})-(\ref{eq_8_21_1}) in Eq.(\ref{V1}), 
we obtain the driving potential as
\begin{eqnarray}
\frac{V_{FF}}{\hbar}=\sumj \frac{v_{FF}(\larger_j,t)}{\hbar},
\label{vff1}
\end{eqnarray}
with
\begin{eqnarray}
\frac{v_{FF}}{\hbar}(\larger_j,t)=-\frac{\pa\alpha}{\pa t}\ep
a_j - \alpha^2\ep^2\frac{\pa a_j}{\pa R}
- \alpha\ep\im \Big{[}\frac{\pa\vap_j}{\pa R}\Big{/}\vap_j\Big{]}\nonumber\\
-\frac{\hbar}{2m}\Big{(} 2\alpha\ep\nab_ja_j\cdot\nab_j\xi_j
+\alpha^2\ep^2(\nab_ja_j)^2 \Big{)}.
\label{vff2}
\end{eqnarray}
In Eq.(\ref{vff2}) $\alpha$, $a_j$, $\vap_j$ and $\xi_j$ abbreviate
$\alpha(t)$, $a(\larger_j,R(\lam))$, $\vap(\larger_j,R(\lam))$ and 
$\xi(\larger_j,R(\lam))$, respectively.
The driving potential in Eq.(\ref{vff1}) is composed of one-body operators.
These results insist that if the ground state of dilute Bose gas is 
represented by Eq.(\ref{phi1}), we can transform the Bose gas 
from a ground state to another corresponding to a different 
adiabatic parameter $R$
in any desired short time.

\section{Conclusion}
\label{Conclusion}
We have presented the acceleration of adiabatic dynamics of quantum systems
composed of identical spinless particles interacting with each other.
We have derived a driving potential, which accelerates the adiabatic 
dynamics, by extending the preceding fast-forward scaling theory formed 
for single particle systems.
The driving potential produces the final state
of an original adiabatic dynamics from the initial state 
of the adiabatic dynamics in any desired short time.
There is no energy excitation at the final time of the manipulation.
Although the driving potential is a many-body potential in general,
it becomes a one-body potential for particular manipulations:
transport of interacting identical particles and control of dilute Bose gas
in the ground state.
We have showed a driving potential for
acceleration of adiabatic transport of interacting particles.
The driving potential, which is linear and time-dependent, 
transports particles without energy excitation. 
It was proven that the driving potential 
conveys not only energy eigenstates but also
general states which is a superposition of energy eigenstates
because the driving potential does not depend on energy levels.

We have also showed an ideal rapid manipulation of a dilute 
Bose gas in the ground state. 
We have derived a formula of the driving potential which accelerates 
adiabatic dynamics of the Bose gas
with the mean field approximation as Eq.(\ref{phi1}).
Naive rapid control of external fields 
without fast-forward theory would deform the states into more complex feature. 
However our results insists that 
we can transform a dilute Bose gas from an initial ground state to another
corresponding to a different external field 
without leaving disturbances in the wave function 
in any short time if the ground state is well represented by the mean field
approximation.

\begin{acknowledgements}
We thank global COE program 
``Weaving Science Web beyond Particle-Matter Hierarchy'' for its 
financial support.
We thank K. Nakamura for useful discussions and comments.
\end{acknowledgements}

\end{document}